# A temporal neural network model for object recognition using a biologically plausible decision making layer

Hamed Heidari-Gorji[1,2], Sajjad Zabbah[2], Reza Ebrahimpour[1,2*]

[1] *Faculty of Computer Engineering, Shahid Rajaee Teacher Training University, P.O. Box: 16785-163, Tehran, Iran.*

[2] *School of Cognitive Sciences, Institute for Research in Fundamental Sciences (IPM), P.O. Box: 19395-5746, Tehran, Iran.*

**Abstract**

Brain can recognize different objects as ones that it has experienced before. The recognition accuracy and its processing time depend on task properties such as viewing condition, level of noise and etc. Recognition accuracy can be well explained by different models. However, less attention has been paid to the processing time and the ones that do, are not biologically plausible. By extracting features temporally as well as utilizing an accumulation to bound decision making model, an object recognition model accounting for both recognition time and accuracy is proposed. To temporally extract informative features in support of possible classes of stimuli, a hierarchical spiking neural network, called spiking HMAX is modified. In the decision making part of the model the extracted information accumulates over time using accumulator units. The input category is determined as soon as any of the accumulators reaches a threshold, called decision bound. Results show that not only does the model follow human accuracy in a psychophysics task better than the classic spiking HMAX model, but also it predicts human response time in each choice. Results provide enough evidence that temporal representation of features are informative since they can improve the accuracy of a biological plausible decision maker over time. This is also in line with the well-known idea of speed accuracy trade-off in decision making studies.

**Keywords:** object recognition; temporal neural network; decision making; speed-accuracy trade-off; Psychophysics; Spiking neural networks

[*] Correspondence: Reza Ebrahimpour, Tel.: +9821 22294035; fax: +9821 22280352. E-mail addresses: Hamed.h@live.com, hamed.heidari@sru.ac.ir (Hamed Heidari-Gorji), s.zabbah@ipm.ir (Sajjad Zabbah), ebrahimpour@ipm.ir, rebrahimpour@sru.ac.ir (Reza Ebrahimpour).

# 1. Introduction

Object categorization is one of the most important high level functions of the brain. Mammals, especially humans are capable to categorize objects with different difficulty levels (controlled by noise, variations etc.) in different times [5, 9]. Object categorization in the brain is thought to be processed in two steps: feature extraction and decision-making [26, 32, 33, 41]. Brain extracts different features over time in the extraction step (from early visual cortex to the inferior temporal cortex) [4, 18, 40] and then the areas involved in decision-making (prefrontal cortex) determine the decision using temporally extracted features [14, 35].

Studies have suggested different biologically plausible models capable to follow aspects of human and monkey behavior such as accuracy [1, 7, 10, 19, 23, 34, 37, 42, 43] or reaction time [24]. One of the first hierarchical cortex-based models for object recognition based on neurophysiological evidence is Neocognitron [8]. In the first layers, Neocognitron neurons are sensitive to simple stimuli such as lines in a specific angle. In the next layers, neurons combine previous layers output and respond to more complex stimuli. HMAX model is another model with similar structure to Neocognitron but with different combination rules. HMAX succeeded in producing a comparable performance to human behavior in an animal vs non-animal task [31]. In fact, modified versions of HMAX have been presented in other studies to explain the mechanisms of object recognition for different tasks [7, 10, 15, 29]. In 2007, Masquelier and Thorpe modified HMAX using spiking neural networks and STDP learning rule [23]. Their model has more biological structure and compared with the original HMAX model, performs better in object categorization.

In all of the abovementioned studies, biologically plausible processes are used to extract informative features from the input stimuli and finally non-biological classifiers such as support vector machine or radial basis function operate on the extracted features to determine the input category. These models can follow human performance in rapid categorization tasks. However, they do not explain the human response time which is an important aspect of response behavior [39] and is in trade off with accuracy (spending more time cause better accuracy) [3, 12]. Although there are some models which explain response time [24], they do not explain the accuracy of choice and the relation between accuracy and reaction time in a categorization task.

Here, we present a computational model based on the spiking HMAX. in addition to accuracy, this model is capable of representing the reaction time and may explain the relationship between these two behavioral characteristics. Features are represented temporally at the feature extraction part of the model and they are transferred to the decision making layer. The decision making layer contains units which accumulate evidence provided by feature extractor layers over time in support of any possible choices. The input category will be recognized as soon as an accumulator reaches a threshold. This accumulation to the bound mechanism of decision making is a well-known biologically plausible decision making model [11, 14, 30]. In addition, it should be noted that temporal representation of the features in the feature extraction layers are informative since they can improve the accuracy in decision making layer over time. Results of the model were compared with accuracy and reaction time of human subjects in a face-house categorization task. The proposed object recognition model closely follows human reaction time and their accuracy better than classic spiking HMAX model.

**2. Materials and Methods**

In the section, we first give details about spiking HMAX model by Masquelier and Thorpe in 2007 as our base model[23]. Next, our proposed model is introduced and we will show how we are using behavioral result of a psychophysics task to adjust model parameters.

**2.1 Base Model**

Spiking HMAX is one of the few object recognition models with simple feedforward architecture which considers time concept. After converging STDP features, this model offers a good tradeoff between selectivity and invariance. Through unsupervised STDP, intermediate-level neurons become selective to the frequent patterns in the presented natural images. Over time, responses get more reliable and faster and their latencies are decreased. This hierarchical network consists of four layers S1, C1, S2, and C2.

**2.1.1 S1 layer**

S1 units correspond to simple cells of the V1 visual cortex as described by Hubel and Wiesel [16] and they detect edges of the presented image in four different orientations. Through application of a convolution operator, S1 units serve to detect the edges in the input image. The convolution matrices were 5*5 in size and approximately were similar to the Gabor filters with a wavelength of 5 and a width of 2. These kernels were in 4 different angles 0+22.5, 45+22.5, 90+22.5, and 135+22.5 in degrees (the added rotation is to prevent focusing on the horizontal and vertical edges). These filters were applied across five different sizes of the input image (100%, 71%, 50%, 35%, and 25%). Overall, S1 output is 4*5=20 images.

**2.1.2 C1 layer**

C1 functionally corresponds to the complex cells of the visual system. In this layer, a 7 * 7 window is max pooled and then slid ahead with an overlap value of unity for each of the 20 images outputted by the S1. The resulting size will be 1/6 of the S1 inputs. For example, for an original S1 input of 454*300, the output is 5 images of the sizes: 75*50, 53*35, 38*25, 26*17, and 19*12 for each orientation angle and the overall number of representation is still 20.

Next in C2 is another max pooling for same size images each belonging to one of the 4 orientations calculated by S1. The max pooling in this stage is done by taking the pixelwise maximum of the 4 m by n images and storing the result in a new m by n matrix.

At this stage in C1, there are 5 different size matrix representations of the original image. Then spike times are calculated as reciprocals of the points within these matrices. The position of the point generating the spike and also the related scales and rotations are stored along with spike times to be utilized in the next layers. By sorting the vectors containing this information according to the spike time, a spike train is generated to be inputted to the next layers. This method for spike generation is supported by empirical evidence of the V1 area of the brain indicating a relation between stimulus contrast and decrease of reaction time.

**2.1.3 S2 layer**

Cells in S2 correspond to the intermediate visual features. In this study, for S2 layer we trained 10 cells for house and 10 cells for face categories. Each cell consists of 5 maps and each map matches in size to one C1 scale. Spikes from C1 affect only the corresponding map with the same scale and the effect matches in size to the S2 receptive fields. If any point within any of the containing maps of a cell reaches the threshold, a spike is sent to C2.

For the training of this layer, we utilized almost the same simple STDP learning method as in the base model. The only difference was that in testing, we made some changes as explained in the following. We decreased the firing threshold for S2 and unlike the base model which has one-winner-take-all mechanism, our model reset the potential of S2 and allows it to reach the threshold for unlimited number of times.

**2.1.4 C2 layer**

Every time, the threshold is hit for S2 map, C2 neuron fires. With the changes we made in the base model, reducing the threshold and allowing the neurons to fire more than once, spike trains were fed to C2 and not just single spikes. In the final stage, we substituted the classifier layer in the base model with a decision making stage.

**2.2 Proposed model**

The proposed model consists of two stages; feature extraction and decision making (fig 1). In feature extraction, a modification of HMAX model called the spiking HMAX model is utilized [23]. The main difference between the spiking model and the original HMAX model is using I&F neurons with a simple STDP training method in spiking HMAX which yields to a temporal representation of features. However, due to utilizing the categorization methods such as support vector machine or radial basis function, the response of the model at the end of simulation is time independent. In other words, the response time is not represented in this model as with other biological recognition models.

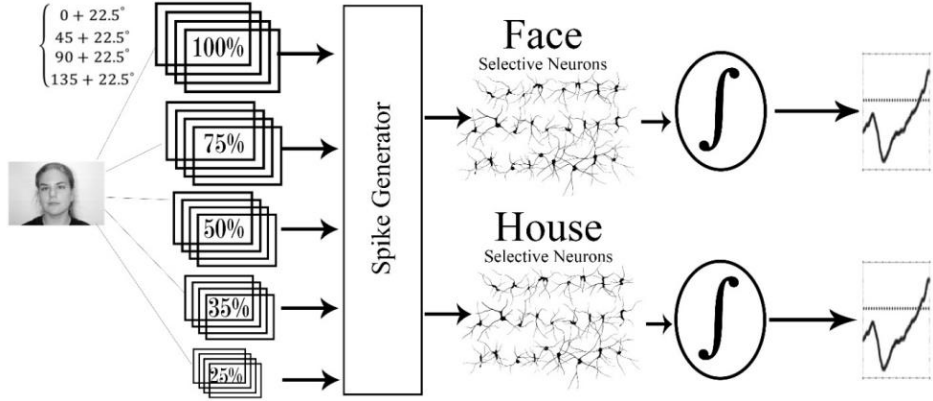

*Figure.1 the image enters the model with different scales and orientations. Then the edge detection and maximization step are performed and the spike data is generated. In the following steps, the spikes are fed into the face selective and house selective neurons. The output of these neurons is connected to the decision making stage. If the neuron reaches the threshold of the face (house) category, images are categorized as face(house)*

The first stage of our model uses the spiking HMAX model as described in the previous section with few modifications. first, we trained two models of spiking HMAX without any changes to the base model with face and house datasets. Afterwards, we used one S1 and one C1 layer from the base model and the output is given to two sets of S2 and C2 cells (the Face Selective Neurons and House Selective Neurons (fig 1)) that were trained with face and house datasets. Unlike the base model where each neuron can only fire one time, we modified the model in a way that neurons are allowed to fire many times (raster plot and average firing rate (AFR) in fig 2).

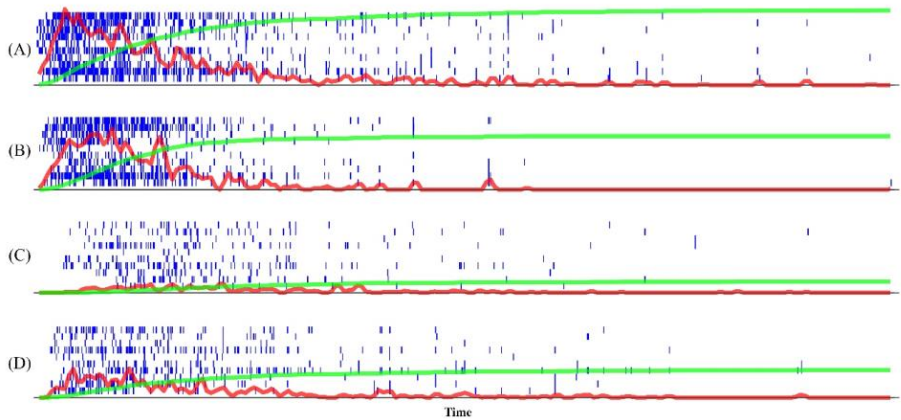

*Figure 2. Model raster plot for 10 face-selective neurons. Red curve indicates running AFR of model neurons. Cumulative AFR is represented by green. (A) face without noise; (B) face with 20% noise; (C) face with 30% noise; (D) house without noise. N.B) neurons are the same STDP face trained neurons across all trials A through D and only the input image is different.*

The last part of the model makes decisions based on the received information from the feature extraction stage. This stage is similar in function to the regions performing decision making such as FEF, LIP, and SC in the brain. The decision making stage in our model takes the input from all last layer neurons of feature extractor stage. This part individually performs information accumulation in favor of each possible class of input image (here face and house) based on the accumulation to threshold model (fig 3 and green curves in fig 2). Therefore, the integrators accumulate the spikes until they hit the threshold and the decision is made. Through training, the decision making model learns to assign a threshold for each of its accumulators.

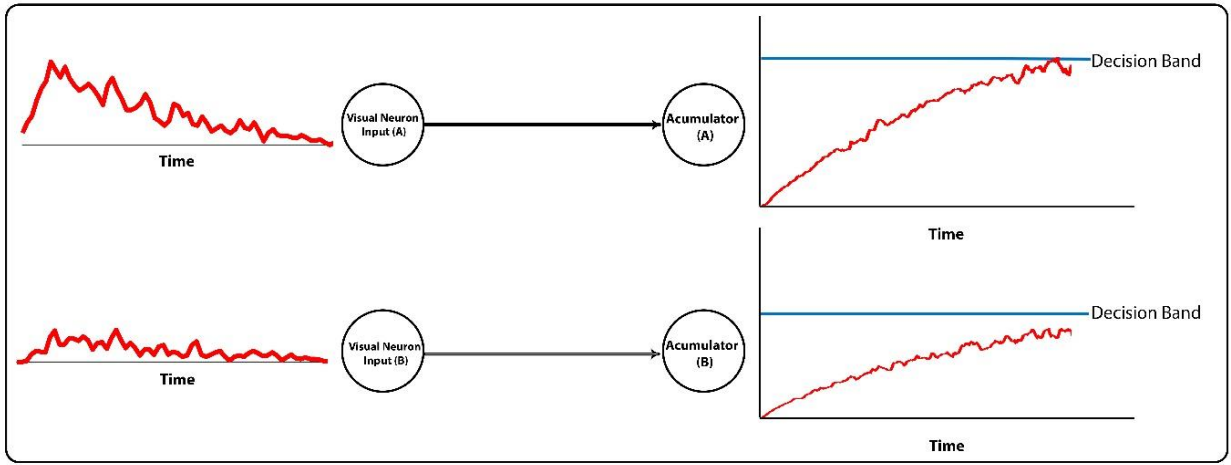

*Figure 3. The schematic of information accumulation region.* *The output firing rate of each feature extractor region is entered to the corresponding accumulator. Information is integrated over time until they reach the threshold and the decision is made. The accumulator that reaches faster to the threshold identifies the class of input image. Decision time is calculated from the start of information accumulation until reaching the threshold.*

**2.2.1 Model components**

In order to solve the recognition problem with two classes of face and house, two different neural populations for house and face processing are considered as shown in figure 1. These two divisions are equivalent to face and house selective regions in the cortex as reported in biological literature [2, 17]. The first stage of the model is simply called feature extraction stage. In other words, the image enters the model and spikes are generated from the input image. Then, the neurons sensitive to face or house receive the previous layer spikes and start their activity. Activity of these population is entered into decision making stage. In fact, the face and house accumulators start to accumulate spikes simultaneously. The accumulator first passing the threshold is the so-called winner and the time spent over accumulation phase in the subthreshold regime is considered as the decision making time. However, for the simulation of human reaction time, the decision time is added to a constant time that corresponds to the motor command transmission time (eq.2). This value is obtained automatically in the training step. It should be mentioned that all the training steps are performed with different images from the training session.

$$MSE = \sum_{i=1}^{10} \sqrt{(RT_{face\_model_i} - RT_{face\_psycho_i})^2 + (RT_{house\_model_i} - RT_{house\_psycho_i})^2} \quad (1)$$

Where $RT_{face\_psycho_i}$ and $RT_{house\_psycho_i}$ represent the average reaction time for the images of house and face in the noise level "i" during the psychophysics test, respectively. To calculate $RT_{X\_model_i}$ ($X = face\ or\ house$) we used the below equation.

$$RT_{X\_model_i} = a \times RT_{X\_model_i} + RT_{motor} \quad (2)$$

Where $RT_{X\_model_i}$ is the reaction time of model x (x = face or house), $\alpha$ is the time scaling parameter, and $RT_{Motor}$ is the required time for motor command transmission. To calculate the threshold in face and house accumulators ($TH_{face}$ and $TH_{house}$ respectively), $RT_{Motor}$ and $\alpha$, the MSE from (1) and (2) is minimized by using Genetic Algorithm. For genetic algorithm we used "ga" function from MATLAB optimization toolbox.

$$RT_{face\_decision} = t\ \ if\ \ AC_{face}(t) == TH_{face}\ and\ AC_{house}(t) < TH_{house} \quad (3)$$

$$RT_{house\_decision} = t\ \ if\ \ AC_{face}(t) < TH_{face}\ and\ AC_{house}(t) == TH_{house} \quad (4)$$

In which, $RT_{motor}$ and $\alpha$ are the same as before, and $RT_{X\_decision}$ is the equivalent decision making time. $RT_{X\_decision}$ is obtained from (3) and (4). $TH_{face}$ and $TH_{house}$ are the threshold of face and house respectively.

$$AC_{face} = \sum_t [v_{face}(t) - u \cdot v_{house}(t)] \quad (5)$$

$$AC_{house} = \sum_t [v_{house}(t) - u \cdot v_{Face}(t)] \quad (6)$$

Where $AC_{face}$ and $AC_{house}$ are integrated information in face and house accumulators obtained from (5) and (6) respectively, $v_{face}(t)$ is the input information value at time t from face selective population (same for house accumulator), and u is a coefficient for the opposing accumulator and was set to zero in our computations.

## 2.3. Psychophysics test

10 subjects (7 men, 3 women, aged between 20-35 years) participated in psychophysics test. All participants were in normal state regarding their visual acuity and had no familiarity with the images used during the test, based on their declaration. The participants were seated on a fixed chair within a constant distance from the monitor in a dark room. The distance from the participants' eyes and the monitor was 59 cm and a chin rest was used to ensure that the distance is maintained throughout the experiment. The display screen was CRT type with a resolution of 600*800 in pixels.

The data set consisted of 200 images with equal numbers (100) for face and house classes. All the images were 300*450 pixels (7x7º) and also gray scaled with 8-bit color code. Each figure was used in 10 different noise levels and in total, 2000 images were generated for the test. Every participant responded to one block of samples with 200 images (100 human face and 100 of house, with 10 images in each level of noise for house or face). Each image was presented only once for each noise level. Participants could see the image only once during the test and there was no chance for getting familiar with the image. The Psychophysics Toolbox.V3 of MATLAB was used to design the task.

**2.3.1. Noise production**

For every image, a 2-D Fourier transform is taken and the phase and magnitude are obtained. The average magnitude of all images is also computed. The noisy phase is a weighted (corresponding to required level of noise) linear combination of the original phase and a phase noise between -$\pi$ and $\pi$. Finally, the noise image is obtained from the inverse Fourier transform of average of magnitude and the noisy phase of main image (fig 4)[28].

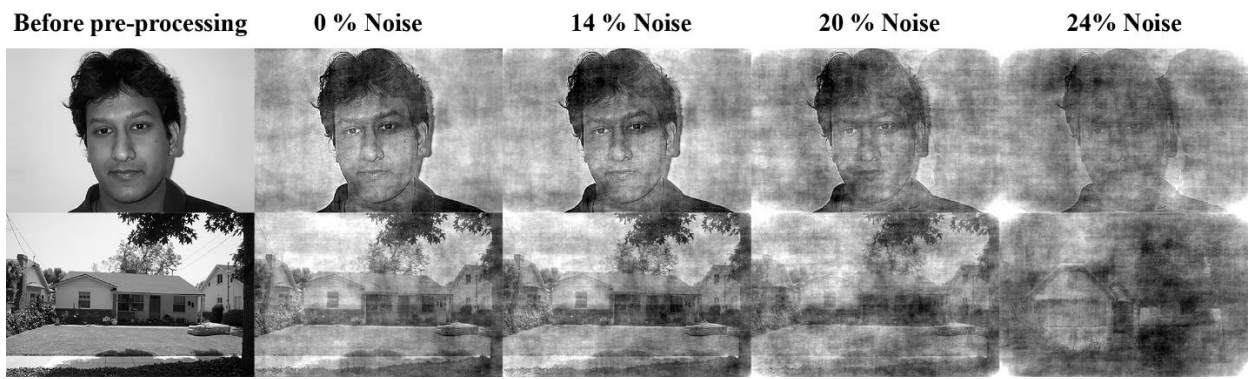

*Figure 4. The sample of used images in the model and Psychophysics test.* The upper row is face and the lower row is house with different noise levels.

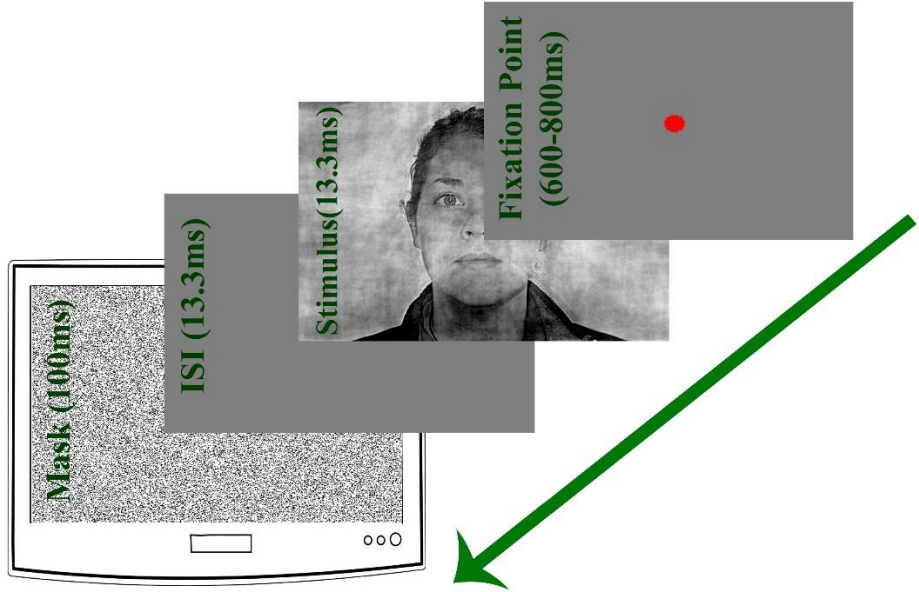

*Figure 5. Psychophysics paradigm. Each trial starts by 600-800 ms presentation of a red dot in the center. The stimulus and a blank gray page will be both presented for 13.3ms sequentially. The trial will be finished by presentation of a noisy mask for 100 ms.*

In all the trials, at the beginning of the test, a fixation point in the center of monitor was shown for a random time between 600-800 milliseconds. In the next step, one image from 200 existing images is shown randomly for 13.3 milliseconds in the monitor. the presented image is a house or a face in one of the 10 noise levels. Then, a gray screen as an Inter stimulus interval (ISI) is shown for 13.3 ms. Afterwards, a noisy mask was presented for 100 ms in order to block the recurrent activities in the brain (fig 5). Subjects report face and house by pressing F and H keys on the keyboard, respectively as fast as accurate they can. The reaction time from this behavioral test is used to set the free parameters of equation (2).

## 3. Results

Our proposed model contains layers corresponding to the visual pathway. Once trained, neurons within these layers become sensitive to the features of house and face in the input image. Depending on the presence and the strength of features in a given input stimulus, neurons in the last layer of the feature extraction part of the model are activated. Face (house) related information represented in face-selective (or house selective) neural activity is accumulated over time by an accumulator in decision making part of the model. Similar to neural decision making in the brain, in this layer, a threshold for the face or house accumulator is obtained. Upon reaching the threshold the model detects a face or house. It should be emphasized that face or house images enter the model without pathway selectivity. Therefore, it is possible that some features in a face image cause activation of house selective neurons which consequently cause the activity of house information accumulator or vice versa. However, each unit of the decision making layer that reaches the threshold determines the model decision and a linear scale of threshold crossing time determines the decision time. To identify model's free parameters, reaction time of human subjects in a behavioral experiment and a genetic algorithm were used. Furthermore, the model performance was compared with the human performance and with another biologically plausible model and some other well-known deep neural networks. Finally, the speed accuracy trade-off that is a well-known behavior of the brain in decision making, was examined in this model.

**3.1 Free parameters adjustment**

Face threshold, house threshold, the model time scale and non-decision time are four free parameters of the model which are adjusted during training of the second stage of the model. The non-decision time which is the motor command time is assumed to be the same for all input images regardless of their difficulty (noise level). These free parameters are estimated through a genetic algorithm so as to minimize the error between model and human reaction times.

**3.2 Model versus human reaction time**

Our results show that the reaction time for either face and house stimuli in a different level of noises are in line with the reaction time of human in different noise level (see figure 6). In this figure, each point is the average of discrimination times for correct decisions for ten different noise levels in the input image. Each noise level includes 20 different images from house and face categories presented to the participants and the model. The light red curve shows the average and the standard error of the participants' reaction time at each noise level. The dark red curve presents the average reaction time of the model in 10 runs. To make the model independent of the training data, the model in each simulation was trained with different images.

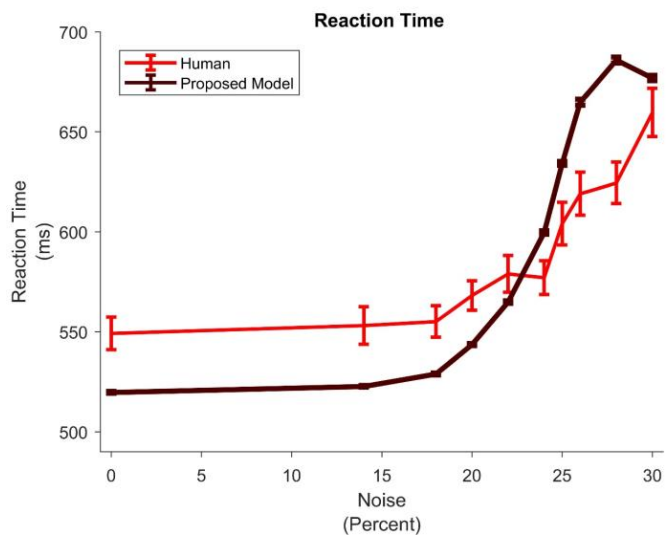

*Figure 6. Human Reaction Time versus Model.* *Light red indicates the human reaction time in a psychophysical experiment and dark red shows the reaction time of the model. The x-axis represents the noise level in percent and the y-axis represents the reaction time in milliseconds. error bars are the standard error.*

The model's behavior which is close to the human behavior in a psychophysics test denotes that a simple structure of information accumulation up to the potential threshold can explain the human behavior in decision making at different times.

**3.3 Model versus human and other models performance**

In contrast to the reaction times, performance in the behavioral tasks was not used in determining the parameters of the model. Therefore, comparing the human and other models is another criterion for validating the model presented in this study. In Figure 7, it can be seen that the model performance is in good agreement with the human performance for different levels of noise. And more importantly, the model that presents information over time to biologically-based decision-making units is much closer to human behavior than the classic spiking HMAX model, which provide the information at a single moment to a non-biologically plausible classifier (RBF).

To compare our results with other models, we used several well-known deep models including VGG16 [36], VGG19 [36], and ResNet50 [13]. For training these nets with house and face data, we used transfer learning method on pretrained models with ImageNet [6]. For this purpose, the last layer of these models with 1000 neurons (number of categories in ImageNet) was replaced by a layer with two neurons (for house and face) and during the training, only weights of the layer just before the last and the last layer were adjusted. Adam optimizer was used for the training.

As expected, deep models achieved almost 100 percent performance on noiseless or low-noise data. The surprising finding was that with increasing levels of noise the performance of deep models relative to human performance was severely dropped. However, by using the temporal information of spikes, the proposed model closely followed the human performance. It is necessary to know that both the deep and proposed models were trained with noiseless data.

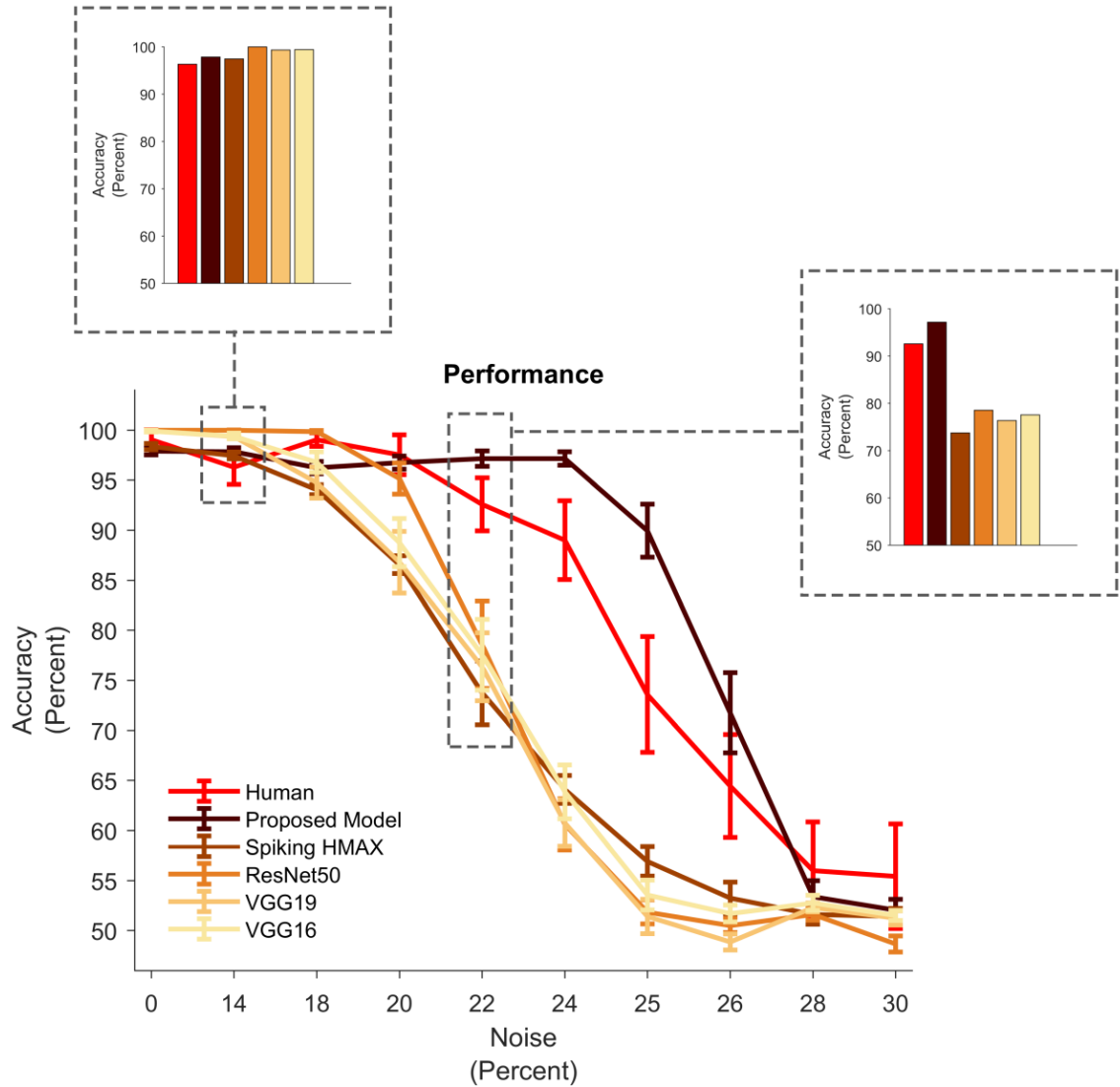

*Figure 7. **Human versus models performance**. For humans, each point represents average performance of 10 people for 20 stimuli in the same noise level (10 face stimulus and 10 house stimulus). For Proposed and spiking HMAX models, each point is the average performance for 10 model results from 10 different trainings of the models. For the deep models (VGG16, VGG19, and ResNet50) transfer learning was used and each point is the average performance for 10 model results from 10 different trainings of the models. Bar graphs emphasize the performance for specific noise levels (14 and 24 percent). The x-axis represents the noise levels and the Y-axis represents the percentage of performance. Error bars are SEM.*

**3.4 How informative are temporal information?**

Masquelier et al, showed that the time of the first spike is sufficient for rapid categorization of simple and noiseless stimulus. They also showed that selectivity to the first spike patterns can be obtained using an STDP learning rule. Thus first spike times contained reliable temporal information in their model. However as shown in figure 7 the first spike cannot categorize noisy stimulus as well as human but using the other spikes the categorization accuracy improves to the range of human accuracy. In order to investigate whether the information provided by the feature extraction part of the model over time is informative or redundant, performance of the model was calculated over time. Figure 8 shows that letting model to use more information over time will increase the categorization performance. In other words, information extracted in different time points are not redundant but they improve the categorization performance. On the other hand, it is noticeable that for the high level noise in the input stimulus, increasing the time of processing does not improve the performance monotonically. Since in these stimuli information is hardly reliable, accumulated noise over time can defect the performance after some point in time.

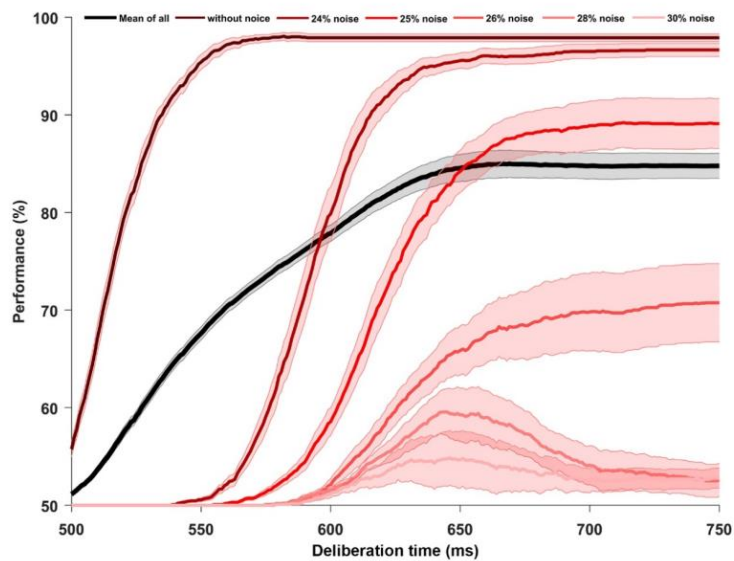

*Figure 8. Model efficiency in limited reaction time and different noise.* X-axis indicates upper boundary for information accumulation (deliberation time). Y-axis shows the performance of the model. The darker the red color, the less the noise level or vice versa, the lighter red color represents higher noise. Black color is the average for all levels of noise.

**3.5 Effects of threshold variation on the performance and the reaction time**

After presenting the compatibility of the model with human behavior in terms of efficiency and reaction time for the face and house classification task, the most important free parameter of the model i.e., the threshold of decision making is evaluated. This parameter generates decision-making trade-off between speed and accuracy, such that, increasing the decision threshold is equivalent to the enhancement of the importance of the decision accuracy against its speed. Therefore, in paradigms in which the accuracy is emphasized, decision threshold is more than the experiments that emphasize the speed of the decision. In this experiment, the efficiency of the decision threshold in the proposed model was examined. Figure 9-A shows the effect of the face threshold variation on the performance (top) and reaction time (bottom) of the model. Different curves stand for three different thresholds in the face accumulator. The optimal state is the threshold obtained by the optimization algorithm using human behavioral data; the two others stand for decreasing the threshold to half the optimal and increasing it to twice the optimal model. Panel B of figure 9 shows the same analysis for house threshold. The x-axis in all parts of figure 9 represents the 200 threshold values, which is considered to be 0.01 times the optimal threshold to twice the desired threshold. It should be noted that, since the correct response for the images at very low thresholds is not enough, there is not any data point at these thresholds in reaction time panels. In addition, at high levels of noise, with the increase in the distance between the two decision bounds, the model can rarely cross each of the thresholds. Therefore, in the figure (9a and 9b below), two primary noise levels have been eliminated.

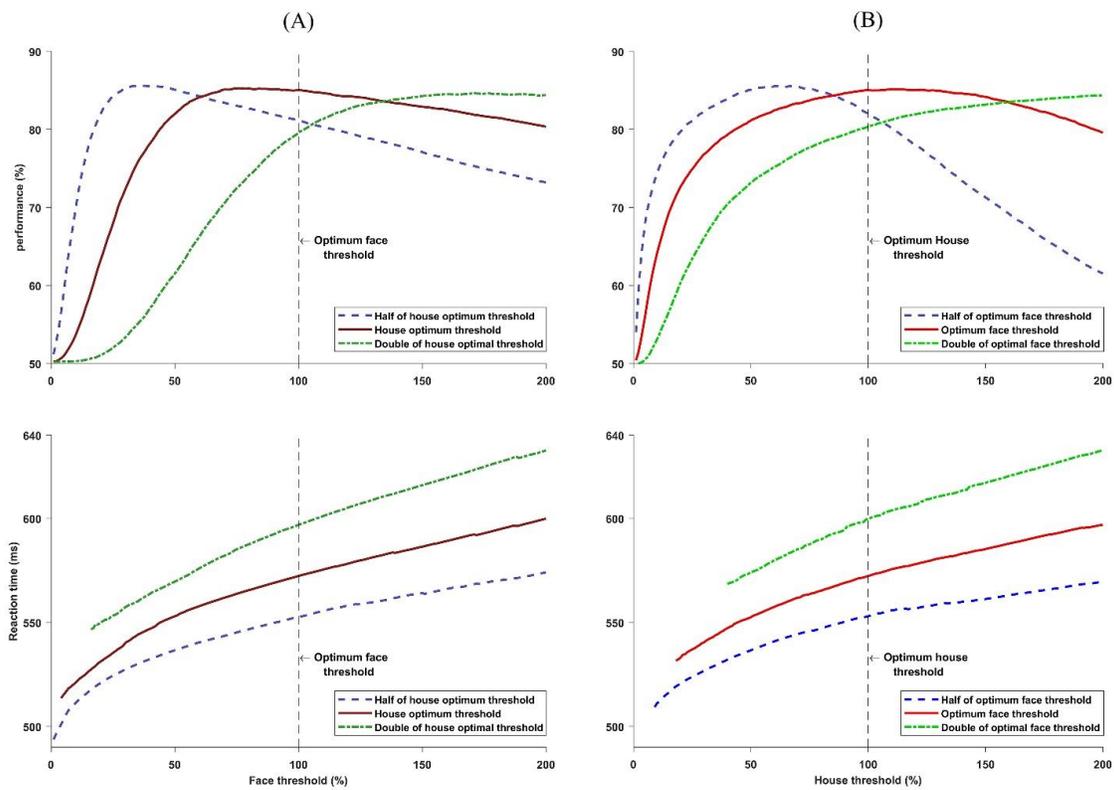

*Figure 9 The effect of the threshold change on the performance and reaction time.* (A-top) The effect of the threshold limit changes on performance. The x-axis shows changes in the model's face threshold and y-axis the change in the model's performance. (A-bottom) The effect of changes in face threshold on reaction time. The x-axis shows changes in the model's face threshold and the y-axis changes the model's reaction time. Each point is the average of the performance of the house and the face images at the specified noise level. (B-top and B-bottom) similar to (A) for changing the threshold of the house. Each point is the average performance of the house and the face images at all levels of noise. The dark red line is the Performance at the optimal House threshold, dark green line is the double threshold performance, and dark blue line, at the threshold that is half the optimal house threshold. The light red lines, light green and blue are the same for the optimal face images.

As shown in figures 9a and b, the increase of the threshold has increased the model efficiency (top panels) and also has led to a decrease in the speed of decision-making (bottom panels) (in the green curve as explicit ascending and in two other curves as a temporary ascending) which is in line with the speed and accuracy trade off in behavior [3, 25]. This result firstly shows that temporal information which is extracted in the first part of the model is meaningful and secondly it shows that the decision threshold plays a role as it does in the behavior.

On the other hand, as shown in Figures. 9a and 9b, the model predicts that, under specific conditions (blue and green curves at high thresholds), the speed accuracy tradeoff is not evident. Blue and green curves in the performance panels do not increase monotonically with thresholds. This result predicts that at some point when distance between two thresholds reaches a specific value speed accuracy tradeoff does not take place in the model behavior. In order to a more accurate observation of this behavior, the figure 9a top, that is the average of ten different noise levels, is represented for each noise level separately in Fig. 10. (The results for the figure (9b top) were the same.)

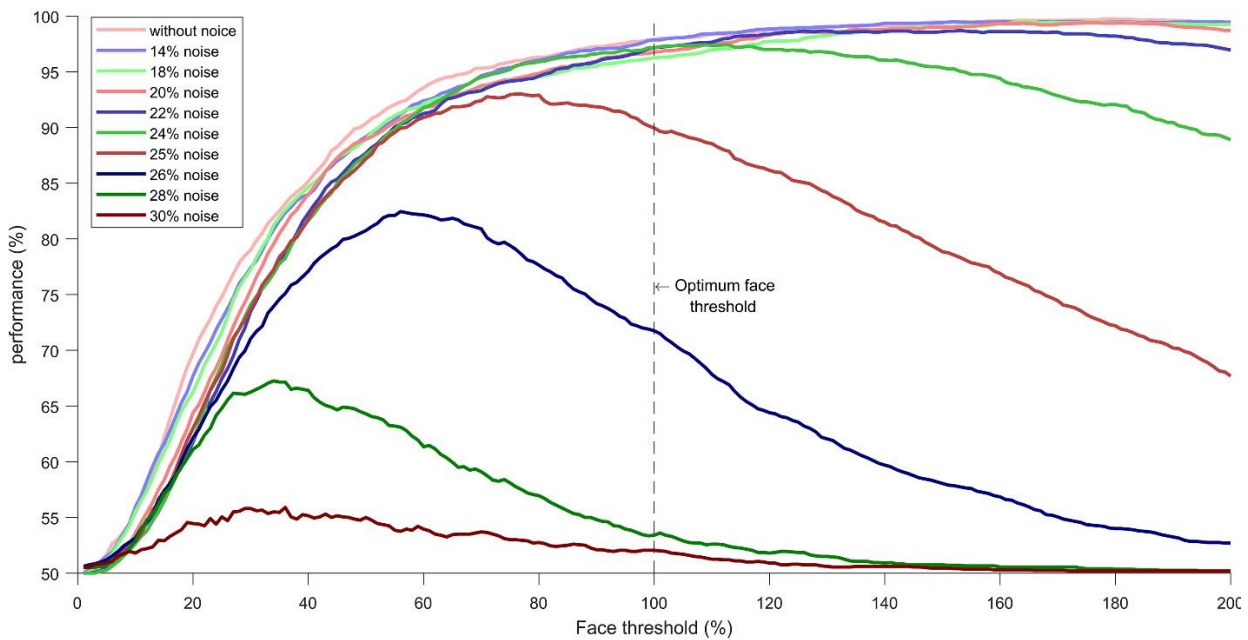

*Figure 10. Performance of the model with threshold variations at different noise levels.* *The x-axis is the face threshold changes in the model and the y-axis is the model performance. Each point is the average performance of 100 house images and 100 face images at the specified noise level.*

Figure 10 shows the performance of the model in discriminating the noise levels in different thresholds of the face. When there is no or less noise in the input image, the performance of the model will always increase by changing the threshold. However, in cases where noise is getting higher, the increase in thresholds initially leads to an increase in the performance and eventually will reduce the model performance. This performance reduction is due to the insufficiency of noisy input stimulus to reach the correct high threshold. Thus, the wrong accumulator which has a lower level of the threshold will be satisfied due to the noise in the stimulus. The results of these two figures are taken into account, the speed accuracy trade-off would be true for a specific range of threshold. This specific range is a function of both input reliability and the distance between thresholds of different accumulators.

## 4. Discussion and Conclusion

Models of object recognition try to evaluate their corresponding brain plausibility by comparing the performance of the model with human performance in psychophysics tasks. However, decision speed, an important characteristic of behavior, is usually ignored in brain plausibility evaluations. In this study, we proposed a plausible model of human object recognition that is able to follow human reaction time as well as its performance in a two forced choice categorization task. In other words, this model is not only able to determine the category of input image but also it determines the time of categorization. The proposed model contains two layers. The first layer is a modified version of a well-known biologically plausible object recognition model (spiking HMAX). This layer extracts perceptual information over time and sends them to the second layer. In the second layer, using a biologically plausible decision making model, two accumulator units, add up the extracted information in support of all possible choices. Decision time is defined as the time of crossing a predefined threshold by each of accumulators. Moreover, decision choice is defined by the accumulator crossing the decision threshold earlier. The accumulation to bound model which we used here as the decision making layer is a well-known model which explain most neurobiological and behavioural findings in the decision making studies[11, 20, 21, 22] . Here, we aimed to propose an object recognition model with a biological plausible decision making layer accounted for response time. Thus, we used accumulation to bound model which has been suggested as a plausible mechanism of making decision in categorizations task [14, 27] and its neural implementation has been well studied as well [38].

Fitting the parameters of the model based on the reaction time of subjects, the model can predict their performance in each stimulus strength. As shown in the result section, without using the information of human performance, the proposed model can follow human performance better than the spiking HMAX. We also showed that speed accuracy trade-off which is a well-known behavior of brain decision process is evident in the model's behavior. This confirms that the representation in the first layer is informative over time. Although, this temporal information is not fully independent (because they come from a single input image), they will improve the performance over time.

In addition, results show that speed accuracy trade-off is not evident in all regimes of input information and decision threshold. As shown in figure 7 the choice would not be necessarily improved by collecting more information. In case that decision threshold is not symmetric for different categories, collecting noisier evidence may result in increasing the probability of satisfying wrong bound. However, this characteristic of decision making, as far as we know, has not been addressed in the psychophysical studies so far.

**Acknowledgements**

We would like to thank Ali Farokhi-nejad, Farzad Shayanfar and Mahboubeh Habibi for their helpful comments and generously editing this manuscript.

**Compliance with Ethical Standards**

**Funding:** This work was supported partially by Cognitive Sciences and Technologies Council under contract number 29602 and Shahid Rajaee Teacher Training University under contract number 330217.

**Conflict of Interest**: The authors declare that they have no conflict of interest.

**Ethical approval**: All experimental procedures were confirmed to guidelines determined by the 1964 Helsinki declaration and were approved by the ethical committee of Shahid Rajaee University. This article does not contain any studies with animals performed by any of the authors.

**Informed consent:** Informed consent was obtained from all individual participants included in the study.